\documentstyle[pra,aps,multicol,eqsecnum,epsfig]{revtex}
 
\def\bo{\begin{figure}}
\def\eo{\end{figure}}

\newcommand\be{\begin{eqnarray}}
\newcommand\ee{\end{eqnarray}}
\newcommand\ba{\begin{array}}
\newcommand\ea{\end{array}}
\def\r{\rangle}
\def\l{\langle}
\def\T{{\rm Tr}}

\begin{document}
\title{
Implementation of quantum maps  by programmable
quantum processors
}
\author{Mark Hillery${}^{1}$,
M\'ario Ziman${}^{2}$
and Vladim\'{\i}r Bu\v{z}ek${}^{2,3}$
}
\address{
${}^{1}$Department of Physics and Astronomy, Hunter College of CUNY, 695,
Park
Avenue, New York, NY 10021, U.S.A.\\
${}^{2}$Research Center for Quantum Information, Slovak Academy of Sciences,
D\'ubravsk\'a cesta 9, 842 28 Bratislava, Slovakia \\
${}^{3}$Faculty of Informatics, Masaryk University, Botanick\'a 68a,
602 00 Brno, Czech Republic
}

\date{5 March 2002}

\maketitle

\begin{abstract}
A quantum processor is a device with a data register and a program
register.  The input to the program register determines the 
operation, which is a completely positive linear map, that will
be performed on the state in the data register.  We develop a
mathematical description for these devices, and apply it to
several different examples of processors.  The problem of finding
a processor that will be able to implement a given set of 
mappings is also examined, and it is shown that while it is
possible to design a finite processor to realize the phase-damping 
channel, it is not possible to do so for the amplitude-damping
channel.  
\end{abstract}

\pacs{PACS Nos. 03.67.-a, 03.67.Lx}

\begin{multicols}{2}

\section{Introduction}
\label{sec1}

The coherent control of individual quantum systems is one of the most
exciting achievements in physics in the last decade 
\cite{Lloyd2001}.
The possibility of controlling quantum dynamics has
far reaching consequences for 
quantum technologies, in particular, for
 quantum computing \cite{Nielsen2000}.
One of the best known applications of coherent control in quantum physics
is the state preparation of  an {\em individual} quantum
system. For example, a particular state of the vibrational motion of a trapped
ion can be prepared by using a well-defined 
sequence of external laser pulses. Another possibility is to focus on
controlling the dynamics, that is, the unitary evolution operator.  One way
of doing this is to realize a particular evolution operator by means of
a sequence of ``elementary'' interactions, which are 
sequentially  turned on and off (for more details see 
 Refs. \cite{Harel1999,Lloyd1999}
and for a specific application to trapped-ions see Ref.~\cite{Hladky2000}
and references therein).

In the theory of quantum coherent control it is  assumed that 
the control  of the dynamics is realized 
via external classical parameters, such as the intensity of a
laser pulse or the duration of an interaction. In this case, the information
that controls the quantum system is classical, and it is set by an 
experimentalist to achieve a single, fixed outcome.  This is analogous
to programming a computer to perform a single task by setting dials
or switches to particular positions, each task requiring different
positions.  

In this paper we will study a different type of quantum control.
We will assume that the information about the  quantum
dynamics of the system under consideration is not represented by 
classical external parameters, but rather is encoded in the
state of another quantum system. A typical example of such an
arrangement is a controlled-NOT (C-NOT) operation (or in general 
a controlled-U operation). In this case, the specific operation performed
on the system, the target, depends on the state of a second quantum system,
the control.  If the control qubit is in the state $|0\rangle$ the
target qubit is left unchanged, but if it is in the state $|1\rangle$,
then a NOT operation is applied to the target qubit.  This means that
this device can perform at least two operations on the target qubit,
the identity and NOT.  There are, however, further possibilities.  Let
us suppose that the control qubit is initially in a superposition of 
$|0\rangle$ and $|1\rangle$, and that we are only interested in the target
qubit at the output of the device, so that we trace out the control qubit
to obtain the reduced density matrix of the target qubit.  The action of
the C-NOT gate on the target qubit can then be described as a completely
positive, linear map acting on the initial density matrix of the target
qubit, with the actual map being determined by the state of the control
qubit.  We take this device to be a model for a programmable quantum gate
array, or quantum processor.  Generally speaking, a  programmable quantum 
processor is a device that implements a completely positive
linear map, which is determined by the state of one quantum system, on a
second quantum system.  
These processors have two registers, the data register and the program
register. The data register contains the quantum system on which the map is
going to be applied, and  the program register contains the quantum system
whose state determines the map.  The third element of this device is a 
fixed array of quantum gates that act on both the program and the data
state.  The virtue of this
arrangement is that we do not have to build a different 
processor every time we want to realize a new map, we simply change the
program state.  This allows us greater flexibility than a device in which 
the map is determined by setting external parameters. For example, it could
be the case that we do not even know what the program state is.  This would
occur when the state of the program register is the output of another
quantum device.  We will refer to the selection of the program state to
perform a desired operation as {\em quantum programming}.

Programmable quantum processors (gate arrays) were first
considered by Nielsen and Chuang \cite{Nielsen1997}.  They
were only interested in the case in which a unitary operation,
rather than a more general completely positive linear map, is performed
on the state in the data register.  If $|\psi\rangle_d$ is the state of
the data register, $|\Xi_{U}\rangle_p$ a program state that implements
the operator $U$ on the data state, and $G$ the overall unitary
operation implemented by the fixed gate array, then their processor carries
out the transformation
\begin{equation}
G(|\psi\rangle_d\otimes |\Xi_{U}\rangle_p )=U|\psi\rangle_d\otimes
|\Xi_{U,\psi}^{\prime}
\rangle_p ,
\label{1.1}
\end{equation}
where $|\Xi_{U,\psi}^{\prime}\rangle_p$ is the state of the program register
after the transformation $G$ has been carried out.  The subscripts
$U$ and $\psi$ indicate that this state can depend on both the operation
$U$ and the state $|\psi\rangle_d$ of the data register $d$.  They were
able to prove a number of results about this device. First, they showed that
the output of the program register does not depend on the data
register, a fact that follows from the unitarity of G.  
Second, they proved that the number of possible programs is equal
to the dimension of the program register.

Let us assume the first of these results and show how to prove the second.
Consider two program states, $|\Xi_{U}\rangle_p$ and $|\Xi_{V}\rangle_p$,
that cause the operators $U$ and $V$, respectively, to act on the
data register.  This implies that
\begin{eqnarray}
G(|\psi\rangle_d\otimes |\Xi_{U}\rangle_p ) & = & U|\psi\rangle_d\otimes
|\Xi^{\prime}_{U}\rangle_p \nonumber \\
G(|\psi\rangle_d\otimes |\Xi_{V}\rangle_p ) & = & V|\psi\rangle_d\otimes
|\Xi^{\prime}_{V}\rangle_p .
\end{eqnarray}
The unitarity of $G$ implies that
\begin{equation}
\label{unig}
_p\langle \Xi_{V}|\Xi_{U}\rangle_p = 
\ _d\langle \psi|V^{-1}U|\psi\rangle_d\,  _p\langle
\Xi_{V}^{\prime}|\Xi_{U}^{\prime}\rangle_p ,
\end{equation}
and if $_p\langle \Xi_{V}^{\prime}|\Xi_{U}^{\prime}\rangle_p \neq 0$, then
\begin{equation}
_d\langle \psi|V^{-1}U|\psi\rangle_d 
=\frac{_p\langle \Xi_{V}|\Xi_{U}\rangle_p}
{_p\langle \Xi_{V}^{\prime}|\Xi_{U}^{\prime}\rangle_p} .
\end{equation}
The left-hand side of this equation depends on $|\psi\rangle_d$ while the
right does not.  The only way this can be true is if
\begin{equation}
V^{-1}U=e^{i\phi} \openone ,
\end{equation}
for some real $\phi$.  This means that the operators $U$ and $V$ are
the same up to a phase.  If we want these operators to be different,
we must have that $_p\langle \Xi_{V}^{\prime}|\Xi_{U}^{\prime}\rangle_p = 0$,
which by Eq. (\ref{unig}) implies that $_p\langle \Xi_{V}|\Xi_{U}\rangle_p =0$.
Therefore, the program states corresponding to different unitary
operators must be orthogonal.  This implies that the dimension of
the program register must be greater than or equal to the number
of different unitary 
operators that can be performed on the data register.

In this paper we would like to consider the more general problem of
quantum processors that realize completely positive linear maps, not
just unitary operators.
The paper  is organized as follows:
In Sec.~\ref{sec2} we derive a formalism for describing and analyzing 
quantum processors, and apply it to both pure and mixed program states.
In Sec.~\ref{sec3} we present several classes of
quantum processors, while Sec.~\ref{sec4} will be devoted to the
problem of the processor ``design'', that is we will discuss specific
processors that are able to implement various 
classes of quantum processes. 
In the
concluding section we will briefly discuss probabilistic quantum processors,
which are based on dynamics conditioned on the results of measurements.

\section{Quantum Processors}
\label{sec2}
A general quantum processor consists of two registers, a data
register and a program register, and a fixed array of quantum
gates.  The input state that goes into the program register encodes
an operation we want to perform on the data register.  We would
first like to show that the action of the processor can be fully
described by a specific set of linear operators.

\subsection{Pure program states}
Let $|\psi\r_d$ be the input state of the data register, $|\Xi
\rangle_{p}$ be the input program state and $G$ be the unitary
operator that describes the action of the array of quantum gates.
If $\{ |j\rangle_{p}|j=1,\ldots N\}$ is a basis for the space of
program states, then we have that
\begin{equation}
G(|\psi\r_d\otimes |\Xi\rangle_{p})=\sum_{j=1}^{N}|j\rangle_{p}
\,_{p}\langle j|G(|\psi\r_d \otimes |\Xi\rangle_{p}) .
\end{equation}
If we define the operator $A_{j}(\Xi )$, which acts on the data register, by
\begin{equation}
A_{j}(\Xi )|\psi\r_d =\,_{p}\langle j|G(|\psi\r_d \otimes |\Xi\rangle_{p}) ,
\end{equation}
then we have that
\begin{equation}
G(|\psi\r_d\otimes |\Xi\rangle_{p})=\sum_{j=1}^{N}A_{j}(\Xi )
|\psi\r_d\otimes |j\rangle_{p} .
\end{equation}
This means that the output density matrix of the data register is
given by
\begin{equation}
\rho^{out}_{d}=\sum_{j=1}^{N}A_{j}(\Xi )|\psi\r_d\ _d\l\psi|
A_{j}^{\dagger}(\Xi ) .
\end{equation}
The operator $A_{j}(\Xi )$ depends on the program state, but it can
be expressed in terms of operators that do not.  Define the operators
\begin{equation}
A_{jk}=A_{j}(|k\rangle ) =\ _p\l j|G|k\r_p,
\label{28}
\end{equation}
where $|k\rangle$ is one of the basis states we have chosen for the
space of program states.  We have that for any program state
$|\Xi\rangle$
\begin{equation}
\label{prgrm}
A_{j}(\Xi )=\sum_{k=1}^{N}\,_{p}\langle k|\Xi\rangle_{p}A_{jk}.
\end{equation}
This means that the operators $A_{jk}$ completely characterize the
processor in the case of pure program states. We shall call these operators
the basis operators for the processor.These operators have the following 
property,
\begin{equation}
\label{orth}
\sum_{j=1}^{N}A_{jk_{1}}^{\dagger}A_{jk_{2}}=
\sum_{j=1}^{N}\l k_1|G^\dagger|j\r\l j|G|k_2\r
=\openone_{d}\delta_{k_{1}k_{2}} ,
\end{equation}
where we have used the decomposition $\sum_j |j\r\l j|=\openone_p$.

An obvious question to ask at this point is whether any set of
operators satisfying Eq.\ (\ref{orth}) corresponds to a quantum
processor.  The following construction allows us to show that this 
is the case \cite{Preskill1998}. Given a set of $N^{2}$ operators
acting on ${\mathcal{H}}_{d}$, we can construct an
operator, $G$, acting on the product space ${\mathcal{H}}_{d}
\otimes{\mathcal{H}}_{p}$, where ${\mathcal{H}}_{p}$ is an
$N$-dimensional space with basis $\{|k\rangle_{p}|k=1,\ldots N\}$.
We set
\begin{equation}
\label{canon}
G=\sum_{j,k=1}^{N}A_{jk}\otimes |j\rangle_{p}\,_{p}\langle k| .
\end{equation}
It is now necessary to verify that $G$ constructed in this way is
unitary.  Noting that
\begin{equation}
\label{adj}
G^{\dagger}=\sum_{j,k=1}^{N}A_{jk}^{\dagger}\otimes 
|k\rangle_{p}\,_{p}\langle j| ,
\end{equation} 
we see that Eq.\ (\ref{orth}) implies that $G^{\dagger}G=\openone$, so
that $G$ preserves the length of vectors and is unitary.

It is possible to express the basis operators for closely related processors
in terms of each other.  For example, if $\{ B_{jk}|j,k=1,\ldots N\}$ are
the basis operators for $G^{\dagger}$, then from Eq.\ (\ref{adj}) we see
that $B_{jk}=A^{\dagger}_{kj}$.  If $G_{1}$ and $G_{2}$ are two processors
(unitary operators) with basis operators $\{ A_{jk}^{(1)}|j,k=1,\ldots N\}$ 
and $\{ A_{jk}^{(2)}|j,k=1,\ldots N\}$, respectively, then the basis operators,
$C_{jk}$, for the processor corresponding to the operator $G_{1}G_{2}$ are
\begin{equation}
\label{prod}
C_{jk}=\sum_{n=1}^{N}A_{jn}^{(1)}A_{nk}^{(2)} .
\end{equation}
This follows immediately if both $G_{1}$ and $G_{2}$ are expressed in the
form given in Eq.\ (\ref{canon}) and then multiplied together.  If we apply
this equation to the case $G_{1}=G$ and $G_{2}=G^{\dagger}$, and note that
$GG^{\dagger}=\openone$, we have that
\begin{equation}
\sum_{j=1}^{N}A_{k_{1}j}A^{\dagger}_{k_{2}j}=\openone_{d} \delta_{k_{1}k_{2}}.
\end{equation}
It is clearly possible to generalize Eq.\ (\ref{prod}) to the case when there 
is a product of more than two operators. 
\subsection{General program states}
Suppose the program is represented by a mixed state
$\varrho_p=\sum_{kl}R_{kl} |k\r\l l|$. Then for the induced mapping
we have
\be
\varrho_d^{out}=\sum_{klmn}R_{kl}A_{mk}\varrho^{in}_d A_{nl}^\dagger
\T_{p}(|m\r_{p}\l n|)
\nonumber
\\
=\sum_{klm}R_{kl}A_{mk}\varrho_d A_{ml}^\dagger.
\ee
We shall denote by ${\cal C}_G$ the set of completely positive linear maps
realizable by using the fixed processor $G$ and any mixed state in the
program space as a program.

Let us now address the question of whether it is possible
to find a second processor, $G^{\prime}$, that can realize any map in
the set ${\cal C}_{G}$ using only pure state programs.  Any mixed state
in ${\cal H}_{p}$ can be purified, but the purification is not unique
\cite{Nielsen2000,Uhlmann1976}.  We begin by defining a new program space, 
${\cal H}_{p^{\prime}}={\cal H}_{p}\otimes {\cal H}_{p}$ and choosing
the purification in the following way
\be
\label{purification}
\varrho_{p}=\sum_{k}\lambda_k |\chi_k\r\l\chi_k |\ \longrightarrow\ \
|\Phi\r_{p^\prime}=\sum_{k}\sqrt{\lambda_k}|\chi_k\r_p\otimes |k\r\, ,
\ee
where $\varrho_p$ has been written in terms of its spectral decomposition. 
We define the unitary operator corresponding to the new processor, which 
acts on the space ${\cal H}_{d}\otimes {\cal H}_{p^{\prime}}$, by 
\be
G^\prime :=G\otimes\openone\, .
\ee
The conjecture is that processor $G^{\prime}$ with the pure program state
$|\Phi\rangle_{p^{\prime}}$ will produce the same mapping as 
the processor $G$ with the mixed program state $\varrho_{p}$.
If this is true, then we will have shown that 
by using only  pure program states
with the processor $G^\prime$, we can implement the entire class of
superoperators ${\cal C}_G$.

In order to prove this we have to show that
\be
\label{4.15}
\T_p G \varrho_d\otimes \varrho_p G^\dagger=
\T_{p^\prime} G^\prime\varrho_d\otimes \varrho_{p^\prime}G^{\prime\dagger}
\ee
for all $\varrho_d$.
The right-hand side of this equation  can be rewritten as
\be
\nonumber
&&
T_{p^\prime} G^\prime\varrho_d\otimes \varrho_{p^\prime}G^{\prime\dagger}
\\
\nonumber
&=&
\T_{p^\prime} \left[\sum_{kl}\sqrt{\lambda_k\lambda_l}\left(
G \varrho_d\otimes|\chi_k\r\l\chi_l|
G^\dagger\right)\otimes|k\r\l l|\right] \\ \nonumber
&=&\sum_{kl}\sqrt{\lambda_k\lambda_l}\T_p\left[\left(
G \varrho_d\otimes|\chi_k\r\l\chi_l|
G^\dagger\right)\delta_{kl}\right]
\\ \nonumber
&=&\T_p\left[
G \varrho_d\otimes\left(\sum_k\lambda_k|\chi_k\r\l\chi_k|\right)G^\dagger
\right]\hspace{1cm}\\ 
&=& \T_p G\varrho_d\otimes \varrho_p G^\dagger \, , 
\ee
which proves Eq.~(\ref{4.15}).
Therefore, we can conclude that it is possible to ``mimic'' mixed program 
states for a given processor by introducing a larger program space 
${\cal H}_{p^\prime}$ and a new processor mapping $G^\prime=G\otimes\openone$.

\subsection{Correspondence between programs and mappings}
We have just seen that two different programs on two different processors
can lead to the same mapping, and now we would like to examime whether
different programs on the same processor can produce identical mappings.
We shall show that this can occur by means of a simple example.  Let
$Q_{p}$ be a projection operator on the program space whose range
has dimension $D$, where $1<D<N$, and let $U_1$ and $U_2$ be two
different unitary operators on the data space.  Consider the processor
given by
\begin{equation}
G=U_{1}\otimes Q_{p}+U_{2}\otimes (\openone_{p}-Q_{p}) .
\end{equation}
Any program state in the range of $Q_{p}$ produces the mapping $U_1$
on the data state, and there are clearly an infinite number of these.
Therefore, we can conclude that there are processors for which many
program states produce the same operation on the data state.

We shall now show that the opposite can also occur, i.e.\ that
there exists a  processor, for which
every program state (mixed or pure) encodes a different superoperator
We will present an example  which illustrates that To do so, we 
utilize results of Ref.~ \cite{Ziman2002} where 
the unitary transformation 
\be
\label{swap}
G=\cos\phi\openone+i\sin\phi S
\ee
was introduced. The swap operator
$S=\sum_{kl}|kl\r\l lk|$ is 
defined in any dimension. The  
so-called {\it partial swap} transformation $G$
acts on two qudits ($d$ dimensional systems).
Let us  restrict our attention to qubits, and identify one of the qubits 
with the data register and other with the  the program system.
In Ref.~\cite{Ziman2002}) it was shown that 
if the program system is prepared in the state $\varrho_p\equiv\xi$,
then the induced map (superoperator) $T_\xi$ is contractive with
its fixed point equal to $\xi$ \cite{blabla3}.
Since each contractive superoperator has only a single
fixed point, we can conclude that different program states
$\xi\ne\xi^\prime$ induce different superoperators , i.e. 
$T_{\xi}\ne T_{\xi^\prime}$. As a result we can conclude that in
the processor given by Eq.\ (\ref{swap}),   
for any value of the parameter $\phi$, the correspondence between
programs and mappings is one-to-one.  Finally, we note that while
here we considered only qubits, the results in this paragraph also
hold for qudits \cite{blabla3}.
\subsection{Equivalent processors}
We shall regard two processors, $G_{1}$ and $G_{2}$ as essentially equivalent
if one can be converted into the other by inserting 
{\em fixed} unitary gate arrays
at the input and output of the program register, that is if
\begin{equation}
\label{eq1}
G_{2}=(\openone_{d}\otimes U_{p1})G_{1}(\openone_{d}\otimes U_{p2}) ,
\end{equation}
where $U_{p1}$ and $U_{p2}$ are unitary transformations on the program space.
If this equation is satisfied, then the processors defined by the two gate 
arrays will perform the same set of operations on data states, but the program
states required to perform a given operation are different, and the outputs 
of the program registers will be different as well.
 If Eq.\ (\ref{eq1}) holds, then
for the basis operators $A_{jk}^{(i)}$ ($i=1,2$) 
associated with the two processors 
we have
\begin{equation}
\label{2.18}
A^{(2)}_{jk} = \sum_{m,n=1}^{N}(U_{p1})_{jm}(U_{p2})_{nk}A^{(1)}_{mn} .
\end{equation}
Therefore, we can regard two processors whose set of operators
$A^{(i)}_{jk}$ are related by the above equation as equivalent \cite{blabla1}.

If processors $1$ and $2$ are equivalent, then they will implement the same
set of superoperators. i.e.\ ${\cal C}_{G_1}={\cal C}_{G_2}$.  In
order to see this, suppose that when the state $|\Xi_{1}\rangle_{p}$
is sent into the program register of processor $1$,  the map $T_{\Xi_1}$, 
with program operators $A^{(1)}_{j}(\Xi_{1})$, is performed on the data
state.  Now consider what happens when send the state $|\Xi_2\rangle_{p}
=U_{p2}^{-1}|\Xi_{1}\rangle_{p}$ into the input of the program register 
of processor $2$.  This will produce the mapping 
$T_{\Xi_{2}}$ on the data state of processor $2$. The relation 
between the program operators $A^{(1)}_{j}(\Xi_{1})$ and 
$A^{(2)}_{j}(\Xi_{2})$ is
\begin{equation}
A^{(2)}_{j}(\Xi_{2})=\sum_{k=1}^{N}(U_{p1})_{jk}A^{(1)}_{1}(\Xi_{1})
\end{equation}
The operators $A^{(1)}_{j}(\Xi_{1})$ are Kraus operators for the mapping
$T_{\Xi_1}$ and the operators $A^{(2)}_{j}(\Xi_{2})$ are Kraus operators 
for the mapping $T_{\Xi_{2}}$.  The above equation implies that the
mappings are identical, $T_{\Xi_1}=T_{\Xi_2}$ \cite{Preskill1998}.
Therefore, any superoperator that can be realized by processor $1$
can be realized by processor $2$.  Similarly, it can be shown that any
superoperator that can be realized by processor $2$ can also be
realized by processor $1$.  This shows that the two processors implement
the same set of superoperators.

A special case of this type of equivalence occurs when the two processors are
simply related by a change of the 
basis in the program space, i.e.\ when $U_{p1}=
U_{p2}^{-1}$.  It is possible to derive  conditions that the basis 
operators of the two processors must satisfy if the processors are to be
equivalent in this more restricted sense.  These 
follow from the fact that the trace is independent of the basis in which it is
taken.  If $U_{p1}=U_{p2}^{-1}$, then ${\rm Tr}_{p}(G_{1})={\rm Tr}_{p}
(G_{2})$, which implies that
\begin{equation}
\sum_{j=1}^{N}A^{(1)}_{jj}=\sum_{j=1}^{N}A^{(2)}_{jj} .
\end{equation}
We also have that ${\rm Tr}_{p}(G_{1}^{n})={\rm Tr}_{p}(G_{2}^{n})$, which for
the case $n=2$ gives us
\begin{equation}
\sum_{j,k=1}^{N}A^{(1)}_{jk}A^{(1)}_{kj}=\sum_{j=1}^{N}A^{(2)}_{jk}
A^{(2)}_{kj}  .
\end{equation}
Clearly, by taking higher values of $n$, we can derive additional
equivalence conditions.

We end this section by summarizing some of our results so far:
\newline
$\bullet$ For a given processor, $G$, any member of the class of all possible 
completely positive linear maps realizable by $G$, ${\cal C}_G$, can be
expressed in terms of the operators $A_{jk}$.
\newline
$\bullet$ We can mimic the action induced by any
mixed program state by a pure program state in a larger program
space.  
\newline
$\bullet$ For any two mappings realized by the processor $G$ and 
the pure state programs $|\Xi_{1}\r_p$ and $|\Xi_{2}\r_p$  the identity
\be
\label{mapcond}
\sum_k A_k^{\dagger}(\Xi_{1}) A_{k}(\Xi_{2})=\l\Xi_{1}|\Xi_{2}\r
\openone_{d} ,
\ee
holds.  This follows directly from Eqs.\ (\ref{prgrm}) and (\ref{orth}).

\section{Classes of processors}
\label{sec3}

In this section we will examine several different kinds of quantum processors.
These will serve to illustrate some of the general considerations in the
previous sections.

{\leftline{\bf 1. U processors}}

Let us suppose that the eigenvectors of the unitary operator,
$G$, that describes the fixed array of gates are tensor products.
In particular, suppose that we have a single orthonormal basis 
for ${\cal H}_{p}$, $\{|k\rangle_{p}|k=1,\ldots N\}$ and a
collection of orthonormal bases for ${\cal H}_{d}$, $\{|\phi_{mk}\rangle_{d}
|k=1,\ldots N, m=1,\ldots M\}$, where $M$ is the dimension of 
${\cal H}_{d}$.  For each value of $k$, the vectors $\{|\phi_{mk}
\rangle_{d}|m=1,\ldots M\}$ form an orthonormal basis for ${\cal H}_{d}$.
We call a processor a {\it U processor} if the eigenvectors of $G$, 
$|\Phi_{mk}\rangle_{dp}$, are of the form
\be
|\Phi_{mk}\r_{dp}=|\phi_{mk}\r_d\otimes|k\r_p.
\ee
In this case the operators $A_{jk}$ are give by $A_{jk}=\delta_{jk}U_j$ 
where $U_{j}$ is unitary (its eigenstates are just $\{|\phi_{mj}
\rangle_{d}|m=1,\ldots M\}$) .  This is the type of processor that was
studied by Chuang and Nielsen \cite{Nielsen1997}, and 
we recall that the dimension of ${\cal H}_p$ is equal to the
number of unitary operators that this type of processor can perform.
The processor acts on the state $|\psi\rangle_{d}\otimes |j\rangle_{p}$
as
\be
\label{unitaryprocessor}
G(|\psi\r_d\otimes|j\r_p)=(U_j|\psi\r_d)\otimes|j\r_p ,
\ee
where $|\psi\r_d$ is an arbitrary data state.

For a general pure program state $|\Xi\r_p=\sum_j\alpha_j|j\r_p$
the encoded mapping, or superoperator, $T_{\Xi}$, is given by the expression
$T_{\Xi}[\varrho_d]=\sum_j |\alpha_j|^2 U_j\varrho_d U_j^\dagger$.
In  the case of a mixed program state
$\varrho_p=\sum_{jk}R_{jk}|j\r\l k|$ the data state is transformed as
$T_{\varrho_{p}}[\varrho_d]=\sum_j R_{jj}U_j\varrho_d U_j^\dagger$. 
Comparing these two
cases we conclude that we can always mimic a mixed program
state by a pure one, in particular, it is enough to set
$\alpha_j=\sqrt{R_{jj}}$. Hence, for this type of processor we can
consider only pure program states without any loss of generality.

Finally, we note that for all program states $|\Xi\r_p$
\be
T_{\Xi}[\frac{1}{d}\openone_{d}]=
\sum_j |\alpha_j|^2 U_j\frac{1}{d}\openone_{d} U_j^\dagger=
\frac{1}{d}\openone_{d} .
\ee
This implies that each element of ${\cal C}_G$ is {\it unital}, i.e.\ it
maps the identity operator into itself.

{\leftline{\bf 2. Y processors}}
A second possibility is to consider a situation that is in some ways the
reverse of the one we just examined.  We have a single orthonormal basis for
${\cal H}_{d}$, $\{ |m\rangle_{d}|m=1,\ldots M\}$, and a set of 
orthonormal bases for ${\cal H}_{p}$, $\{ |\chi_{mk}\rangle_{p}|k=1,\ldots 
N\}$, where $m=1,\ldots M$ labels the bases and the index $k$ labels the
individual basis elements.  We again assume that the eigenvectors of $G$,
$|\Phi_{mk}\rangle_{dp}$ are tensor products, but now they are given by  
\be
|\Phi_{mk}\r_{dp}=|m\r_d\otimes|\chi_{mk}\r_p.
\ee
In this case the processor can be expressed as $G=\sum_m |m\r_d\l m|
\otimes U_m$, where $U_{m}$ is unitary and has eigenvectors
$\{ |\chi_{mk}\rangle_{p}|k=1,\ldots N\}$.  We find the operators $A_{jk}$
by first choosing a single orthonormal basis in  ${\cal H}_p$,
$\{|k\r_p\}$, and computing 
\be
\nonumber
A_{jk}&=& _p\l j|G|k\r_p=\sum_m |m\r\l m|\l j|U_m|k\r\\
&=&
\sum_m (U_m)_{jk}|m\r\l m|.
\ee
The maps produced by $Y$ processors are unital, as can be seen from 
\be
\nonumber
\sum_j A_{jk_1}A^\dagger_{jk_2}&=&\sum_j\sum_{ab} (U_m)_{jk_1}
(U^\dagger_n)_{jk_2}|m\r\l m|n\r\l n|\\
\nonumber &=&
\sum_{ja}(U_m)_{jk_1}(U_m^\dagger)_{jk_2}|m\r\l m|\\
&=&
\delta_{k_1 k_2}\sum_m |m\r\l m|=\delta_{k_1k_2}\openone
\ee

The action of a $Y$ processor is particularly simple if all of the operators
$U_m$ have some common eigenstates, and the program state is one of them.
Suppose that $U_{m}|\Xi\rangle_{p}=e^{i\phi_{m}}|\Xi\rangle_{p}$, then
\begin{equation}
G(\sum_{m} c_m |m\rangle_{d}\otimes |\Xi\rangle_{p})=(\sum_{m}c_m
e^{i\phi_{m}}|m\rangle_{d})\otimes |\Xi\rangle_{p} .
\end{equation}
In summary, we can say that both the $U$ and $Y$ processors are controlled-U
gates; in the $U$ processor, the control system is the program and the target
is the data, and in the $Y$ processor, it is the target that is the program 
and the control that is the data.

\leftline{\bf 3. $\bf U^\prime$  processors}
Let us consider 
a simple modification of the $U$ processor, which we shall call  the 
$U^\prime$ processor.  Suppose we have two different orthonormal bases
of ${\cal H}_{p}$, $\{ |k\rangle_{p}\}$ and $\{ |\chi_{k}\rangle_{p}\}$. 
We define a $U^{\prime}$ processor to have a unitary operator, $G$, of
the form
\be
\label{n1}
G=\sum_k U_k \otimes |k\r_p\l \chi_k | .
\ee
This looks like a new kind of processor, but it is actually equivalent to a
$U$ processor.  This can be seen immediately if we realize that there 
exists a unitary operator, $U_{p}$, acting on ${\cal H}_{p}$ such that
$|\chi_{k}\rangle_{p}=U_{p}|k\rangle_{p}$.  Therefore, we have that
\begin{equation}
G=(\sum_{k}|k\rangle_{p}\langle k|)(\openone_{d}\otimes U_{p}^{\dagger}),
\end{equation}
so that $G$ is, in fact, equivalent to a $U$ processor.

\leftline{\bf 4. $\bf Y^\prime$ processors}
Now let us try a modification of the $Y$ processor in the same spirit
as the one we just made to the $U$ processor.
Suppose we have two different orthonormal bases
of ${\cal H}_{d}$, $\{ |m\rangle_{p}\}$ and $\{ |\phi_{m}\rangle_{d}\}$. 
We define a $Y^{\prime}$ processor to have a unitary operator, $G$, of
the form
\be
G=\sum_m |m\r_d\l \phi_{m}|\otimes U_m .
\ee
For the operators $A_{jk}$ we obtain
\be
A_{jk}=\,_{p}\l j|G|k\r_{p}=\sum_m |m\r\l \phi_{m}|(U_m)_{jk}.
\ee
This type of processor is not equivalent to a $Y$ processor.  It
does, however, share the property of producing unital maps as can
be seen from
\be
\nonumber
\sum_j A_{jk_1}A^\dagger_{jk_2}&=&
\sum_{j,m,n} |m\r\l \phi_{m}|\phi_{n}\r\l n|(U_m)_{jk_1}
(U^\dagger_n)_{k_2j}\\
\nonumber
&=& \sum_{j,m} |m\r\l m|(U^\dagger_m)_{k_2 j}(U_m)_{jk_1}=
\delta_{k_1 k_2}\sum_m |m\r\l m|\\
&=&\delta_{k_1 k_2}\openone_d ,
\ee
which implies that for any program state, the identity on ${\cal H}_{d}$
is mapped into itself. 

{\leftline{\bf 5. Covariant processors}}
Another class of processors that may be of interest are {\em covariant}
processors.  Covariance has proven to be an important property in the
study of quantum machines.  Covariant processors 
have the property that if the processor maps
the input data state $\varrho_{in}=|\psi\r_d\ _d\l\psi|$,
which we shall assume is a
qudit, onto the output density matrix
$\rho_{out}$, then it maps the input state $U|\psi\r_d$ onto
the output density matrix $U\rho_{out}U^{-1}$, for all
$U\in \mathcal{G}$, where $\mathcal{G}$ is a subgroup of
$SU(D)$, for some subset $\mathcal{S}$ of all possible program states 
\cite{blabla2}.
This relation implies that if 
$|\Xi\rangle\in\mathcal{S}$, then the operators
$A_{j}(\Xi )$ satisfy the relation
\begin{equation}
\label{covariant}
\sum_{j=1}^{N}UA_{j}(\Xi )\varrho_{in}A_{j}^{\dagger}
(\Xi )U^{-1} = \sum_{j=1}^{N}A_{j}(\Xi )U\varrho_{in}
U^{-1}A_{j}^{\dagger}(\Xi ) ,
\end{equation}
for all $U\in\mathcal{G}$.  Let us now consider the case ${\mathcal{G}}=SU(D)$.
If we take $\rho_{in}$ to be $\openone_{d}/d$, we find 
\begin{equation}
\sum_{j=1}^{N}UA_{j}(\Xi )A_{j}^{\dagger}(\Xi )
U^{-1} = \sum_{j=1}^{N}A_{j}(\Xi )A_{j}^{\dagger}(\Xi ) .
\end{equation}
Because this holds for all $U\in SU(D)$, Schur's Lemma implies
that
\begin{equation}
\label{5.3}
\sum_{j=1}^{N}A_{j}(\Xi )A_{j}^{\dagger}(\Xi )=c\, \openone ,
\end{equation}
where $c$ is a constant.  Taking the trace of both sides of Eq.(\ref{5.3})
 we find
\begin{equation}
{\rm Tr}\left(\sum_{j=1}^{N}A_{j}(\Xi )A_{j}^{\dagger}(\Xi )\right)
N=c\,{\rm Tr}(\openone)=c\, N ,
\end{equation}
so that $c=1$.  Because this relation holds for any program state,
we have that
\begin{equation}
\label{cov}
\sum_{j=1}^{N}A_{jk_{1}}A^{\dagger}_{jk_{2}}=\delta_{k_{1}k_{2}}
\openone_{d} ,
\end{equation}
which implies that the maps produced by a processor that is covariant
with respect to $SU(D)$ are unital.

Let us briefly consider an example in order to show that a nontrivial
covariant processor with respect to $SU(2)$ exists. We shall examine
a processor provided by the quantum information distributor
\cite{Braunstein2000}.
The program state of this device consists of two qubits and the data state is
one qubit.  The unitary operator, $G$ can be implemented by a sequence of four
controlled-NOT gates.  A controlled-NOT gate acting on qubits $j$ and $k$,  
where $j$ is the control bit and $k$ is the target bit, is described by the
operator
\begin{equation}
D_{jk}|m\r_{j}|n\r_{k}=|m\r_{j}|m\oplus n\r_{k}  ,
\end{equation}
where $m,n=0$ or $1$, and the addition is modulo $2$.  If we denote the data
qubit as qubit $1$ and the two program qubits as qubits $2$ and $3$, then
the operator $G$ for this processor is
\begin{equation}
G=D_{31}D_{21}D_{13}D_{12} .
\end{equation}

For the set of program states, ${\cal S}$, we shall consider two-qubit
states of the form
\begin{equation}
\label{alphbet}
|\Xi\r = \alpha |\Xi_{00}\r_{23} + \beta |\Phi\r_{23} ,
\end{equation}
where 
\begin{eqnarray}
|\Xi_{00}\r & = & \frac{1}{\sqrt{2}}(|0\r_{2}|0\r_{3}+|1\r_{2}|1\r_{3})
\nonumber \\
|\Phi\r & = & \frac{1}{\sqrt{2}}|0\r_{2}(|0\r_{3}+|1\r_{3}) ,
\end{eqnarray}
and $\alpha$ and $\beta$ are real, and $\alpha^{2}+\beta^{2}+\alpha\beta =1$.
If the data register at the input is described by the state
$\varrho_{in}$, then at the output of the processor we find
the data register in the state
\begin{equation}
\rho_{out}=(1-\beta^{2})\varrho_{in} +\frac{\beta^{2}}{2}\openone_{d}
\, .
\end{equation}
The action of this processor is clearly covariant with respect to any 
transformation in $SU(2)$.

\section{Processor design}
\label{sec4}
In the previous sections we have studied sets of superoperators that 
a given processor can perform.  We would now like to turn the problem around 
and suppose that we have a given set of superoperators, and our aim is to
construct a processor that will be able to execute them. 
We already know that it is impossible to find a processor that will perform
all  superoperators.  In particular, if the set of superoperators we are trying
to implement contains an uncountable set of unitary superoperators, then the
set of superoperators cannot be performed by a single processor.

Here we will ask more modest question: Under what circumstances we are  able
to find a processor that will perform some one-parameter set of
superoperators?  In particular, suppose
that we have the superoperators $T_{\theta}$, where the parameter 
$\theta$ varies 
over some range, and that these operators have a Kraus representation 
$\{ B_{j}(\theta)|j=1,\ldots M\}$ such that
\begin{equation}
T_{\theta}[\rho ]= \sum_{j=1}^{M}B_{j}(\theta )\rho B^{\dagger}_{j}
(\theta ) .
\end{equation}
Our aim is to find a unitary operator, $G$, and a set of program states 
$|\Xi (\theta )\r_p$ so that
\begin{equation} 
T_{\theta}[\rho_{d}] = G(\rho_{d}\otimes |\Xi (\theta )\r_{p}\,_{p}\l\Xi
(\theta )|)G^{\dagger} .
\end{equation}

The operators $A_{j}(\Xi )$ that represent the action of the processor on the
data states when the program state is $|\Xi\r$, 
are now functions of $\theta$ and
we shall denote them as $A_j(\theta )$.  Our processor then 
transforms the input data state $\rho_{d}$ into the output 
state, $\rho_{d}^{(out)}$
\begin{equation}
\rho_{d}^{(out)}=\sum_{j=1}^{N}A_{j}(\theta )\rho_{d}A_{j}^{\dagger}(\theta ).
\end{equation}
We note that the operators $\{ A_{j}(\theta )|j=1,\ldots N\}$ also constitute  
a Kraus representation of the superoperator $T_{\theta}$.  The Kraus
representation of a superoperator is not unique; any two different Kraus
representations of the same superoperator, $\{ B_{j}|j=1,\ldots M\}$ and 
$\{ C_{j}|j=1,\ldots N\}$, where $N\ge M$, are related as follows,
\cite{Preskill1998}
\begin{equation}
\label{bla1}
C_{j}=\sum_{k=1}^{N}U_{kj}B_{k} ,
\end{equation}
where $U_{kj}$ is a unitary matrix. It is understood that if $N>M$, then
zero operators are added to the set $\{ B_{j}|j=1,\ldots M\}$ so that the 
two sets of operators have the same cardinality.

In what follows we will study two single-qubit quantum channels, the 
phase-damping channel and the amplitude-damping channel. We will show that the
former can be realized by a finite quantum processor, while the second
cannot.

\leftline{\bf 1. Phase-damping channel.}
The phase-damping channel is described by the
map $T_{\theta}$ that is determined by the Kraus
operators  $B_1(\theta )=\sqrt{\theta}\openone$
and $B_2(\theta )=\sqrt{1-\theta}\sigma_z$, where both 
$\sigma_z$ and $\openone$ are unitary operators, and 
$0\le \theta \le 1$ \cite{Holevo1982,Nielsen2000}.
Hence for the phase-damping map we find
\be
T_{\theta}[\varrho_d]=\theta\openone\varrho_d\openone+
(1-\theta )\sigma_z\varrho_d\sigma_z^\dagger,
\ee
where $\varrho_{d}$ is the input qubit state.
We can design  the corresponding 
processor using Eq.~(\ref{unitaryprocessor}), that is 
\be
G^{phase}|\phi\r_d\otimes|k\r_p=(U_k|\phi\r_d)\otimes|k\r_p\, ,
\ee
where $k=1,2$ and $U_1=\openone,U_2=\sigma_z$. The program state in which
the required transformation $T_{\theta}$ is encoded is given by
$|\Xi(\theta )\r_p=\sqrt{\theta}|0\r_{p}+\sqrt{1-\theta}|1\r_p$.
Note that in this case the program operators, $A_{j}(\theta )$, for
$j=1,2$, are equal to the corresponding Kraus operators, i.e.\ $A_{j}
(\theta )=B_{j}(\theta )$.
Therefore, we can execute the entire
one parameter set of superoperators $T_{\theta}$ merely by changing the 
program state we send into the processor, and the dimension of the
program space is two.

\leftline{\bf 2. Amplitude damping channel.} 
The amplitude-damping map $S_{\theta}$ is given by the Kraus operators
$B_1(\theta )=|0\r\l 0|+\sqrt{1-\theta}|1\r\l 1|$ and 
$B_2(\theta )=\sqrt{\theta}|0\r\l 1|$, where again, $0 \le \theta
\le 1$.  In designing a processor to realize this channel, we would
again like to assume that the program operators are the same as the
Kraus operators, $B_{1}(\theta )$ and $B_{2}(\theta )$.  In this case,
however, we have a problem.  The program operators must satisfy Eq.\ 
(\ref{mapcond}), but
\begin{eqnarray}
\sum_{j=1}^{2}B_{j}^{\dagger}(\theta_{1})B_{j}(\theta_{2})
\nonumber \\
=|0\rangle
\langle 0|+(\sqrt{\theta_{1}\theta_{2}}+\sqrt{(1-\theta_{1})(1-
\theta_{2})})|1\rangle\langle 1| ,
\end{eqnarray}
and the right-hand side of this equation is not, in general, proportional
to the identity.

What we now must do is try to find a Kraus representation for this channel
that does satisfy Eq.\ (\ref{mapcond}).  In particular, we assume that
\begin{equation}
C_{k}(\theta )=\sum_{k=1}^{N}U_{kj}(\theta )B_{j}(\theta ) ,
\end{equation}
where $U(\theta )$ is an $N\times N$ unitary matrix, and $B_{j}
(\theta )=0$ for $j>2$.  In addition, we want 
\begin{equation}
\label{krauscon}
\sum_{j=1}^{N}C_{j}^{\dagger}(\theta_{1})C_{j}(\theta_{2})=f(\theta_{1},
\theta_{2})\openone ,
\end{equation}
where $f(\theta_{1},\theta_{2})$ is a function whose magnitude is less
that or equal to one. The operators $C_{j}(\theta)$ would then be
candidates for the program operators, $A_{j}(\theta )$.  What we
will show is that there is no Kraus representation with $N$ finite
that satisfies these conditions.  Because the number of program
operators is equal to the dimension of the program space, this will
show that there is no finite quantum processor that can realize the
family of superoperators that describes the amplitude-damping channel.

If Eq.\ (\ref{krauscon}) is to hold, then the coefficients of $|0\rangle
\langle 0|$ and $|1\rangle\langle 1|$ must be the same.  Inserting the 
explicit expressions for $C_{j}(\theta )$ in terms of $B_{1}(\theta )$
and $B_{2}(\theta )$, this condition becomes
\begin{eqnarray}
(1-\sqrt{(1-\theta_{1})(1-\theta_{2})})\sum_{j=1}^{N}U_{1j}^{\ast}
(\theta_{1})U_{1j}(\theta_{2}) \nonumber \\
=\sqrt{\theta_{1}\theta_{2}}\sum_{j=1}^{N}
U_{2j}^{\ast}(\theta_{1})U_{2j}(\theta_{2}).
\end{eqnarray}
We can now make use of the fact that the rows of a unitary matrix 
constitute orthonormal vectors and the Schwarz inequality to show that
the magnitude of the sum on the right-hand side of this equation is
less than or equal to one.  This give us that
\begin{equation}
\label{krauscon2}
|\sum_{j=1}^{N}U_{1j}^{\ast}(\theta_{1})U_{1j}(\theta_{2})|\leq
\frac{\sqrt{\theta_{1}\theta_{2}}}{1-\sqrt{(1-\theta_{1})(1-
\theta_{2})}}.
\end{equation}

We now need the result that if $\{ v_{j}|j=1,\ldots N\}$ are vectors of
length $1$, and $|\langle v_{j}|v_{k}\rangle |<1/(N-1)$, then $\{ v_{j}
|j=1,\ldots N\}$ are linearly independent \cite{Vidal2000}.  The proof
is quite short, so we give it here.  If the vectors are linearly 
dependent, then there are constants $c_{j}$, at least some of which
are not zero, such that
\begin{equation}
\sum_{j=1}^{N}c_{j}|v_{j}\rangle =0 .
\end{equation}
Taking the inner product of both sides with $|v_{k}\rangle$ we find that
\begin{eqnarray}
|c_{k}|& =& |\sum_{j\neq k}c_{j}\langle v_{k}|v_{j}\rangle | \nonumber \\
 & < & \frac{1}{N-1} \sum_{j\neq k}|c_{j}| .
\end{eqnarray}
Summing both sides of the above inequality over $k$ gives us that
\begin{equation}
\sum_{k=1}^{N}|c_{k}|< \frac{1}{N-1}\sum_{k=1}^{N}\sum_{j\neq k}|c_{j}|
=\sum_{k=1}^{N}|c_{k}| ,
\end{equation}
which is clearly impossible.  Therefore, the vectors must be linearly
independent.

This can now be applied to the first row of the unitary matrix $U(\theta )$,
which we can think of as an $N$-component normalized vector, 
which we shall call $u_{0}(\theta )$.  What we
will show is that we can find arbitrarily many of these vectors whose
inner products can be made arbitrarily small.  The result in the previous
paragraph then implies that these vectors are linearly independent, but
this contradicts the fact that they lie in an $N$-dimensional space.  Hence,
there must be an infinite number of Kraus operators, and the program space
must be infinite dimensional.

In order to study the inner products of the vectors $u_{0}(\theta )$
for different values of $\theta$, we need to examine the function 
appearing on the right-hand side of Eq.\ (\ref{krauscon2})  
\begin{equation}
g(\theta_{1},\theta_{2})=\frac{\sqrt{\theta_{1}\theta_{2}}}
{1-\sqrt{(1-\theta_{1})(1-\theta_{2})}}.
\end{equation}
Using the fact that if $0\leq \theta \leq 1$, then $\sqrt{1-\theta}\leq 
1-(\theta /2)$, we have that for $0\leq \theta_{j} \leq 1$, $j=1,2$
\begin{equation}
g(\theta_{1},\theta_{2})\leq \frac{2\sqrt{\theta_{1}\theta_{2}}}
{\theta_{1}+\theta_{2}-(\theta_{1}\theta_{2}/2)} .
\end{equation}
Finally, noting that for $\theta_{1}$ and $\theta_{2}$ between $0$ and
$1$, 
\begin{equation}
\frac{\theta_{1}+\theta_{2}}{\theta_{1}+\theta_{2}-(\theta_{1}\theta_{2}/2)}
\leq \frac{4}{3} ,
\end{equation}
we see that
\begin{equation}
g(\theta_{1},\theta_{2})\leq \frac{8\sqrt{\theta_{1}\theta_{2}}}
{3(\theta_{1}+\theta_{2})} .
\end{equation} 

We can make use of this bound, if we choose, for any positive integer
$M$, the sequence $\zeta_{n}=[1/(16M^{2})]^{n}$, where $n=1,\ldots$.  
If $\theta_{1}=\zeta_{n}$ and $\theta_{2}=\zeta_{m}$ where $m>n$, then
\begin{equation}
g(\theta_{1},\theta_{2})\leq \frac{8}{3}\frac{1}{(4M)^{m-n}}.
\end{equation}
The vectors $\{ u_{0}(\zeta_{m})|m=1,\ldots M\}$ have
pairwise inner products whose magnitudes are less than $1/M$, and,
therefore, they are linearly independent.  As these vectors have
$N$ components, if we choose $M>N$ we have a contradiction.  This,
as we stated before, implies that the number of Kraus operators is
infinite, and that the amplitude-damping channel cannot be realized 
by a finite quantum processor.

\section{Conclusion}
In this paper we have presented a theory of programmable quantum
processors that allows us to realize completely positive maps on
quantum systems.  We have introduced several
classes of quantum processors and have discussed the design of 
processors to realize particular classes of superoperators.
In our discussion we focused on
the situation when no measurements are performed on the 
program register.

In concluding this paper let us briefly comment on the fact
that if  we allow dynamics conditioned on the results of 
measurements on the program register, new classes of maps can be realized. 
One version of quantum processors with conditional dynamics,
whose  operating
principle is that of quantum teleportation, was discussed by
Nielsen and Chuang \cite{Nielsen2000}.  Here we shall present a different
example. Consider a  processor consisting or a single C-NOT gate in
which the program register consists of the control qubit, and the data
register consists of the data qubit. If the program qubit is initially
in the state
\begin{equation}
|\Xi\rangle_{p}=\alpha |0\rangle + \beta |1\rangle ,
\end{equation}
and the data qubit in the state $|\psi\rangle_{d}$, 
then the output of our simple processor is the state
\begin{equation}
|\Phi_{out}\rangle_{dp}=\alpha |\psi\rangle_{d}|0\rangle_{p}+
\beta\sigma_{x}|\psi\rangle_{d}|1\rangle_{p} .
\end{equation}
If we trace out the program register we obtain the output density
matrix
\begin{equation}
\rho_{out}^{(1)}=|\alpha |^{2}\rho_{in}+|\beta |^{2}\sigma_{x}\rho_{in}
\sigma_{x} ,
\end{equation}
where $\rho_{in}=|\psi\rangle_{d}\langle\psi |$.
If, on the other hand, we measure the output of the program register in 
the $|\pm x\rangle$ basis, where
\begin{equation}
|\pm x\rangle =\frac{1}{\sqrt{2}}(|0\rangle \pm |1\rangle ).
\end{equation}
and only accept the output of the data register if we get $|+x\rangle$,
then we find for the output state of the data register
\begin{equation}
\rho_{out}^{(2)}=K(\alpha\openone +\beta\sigma_{x})\rho_{in}
(\alpha^{\ast}\openone +\beta^{\ast}\sigma_{x}) ,
\end{equation}
where $K$ is a normalization constant.  We note that the sets of
mappings realized by the two different procedures are not the same.

While the addition of conditional measurements to quantum processors
allows us to realize a different set of mappings, there is, however,
 a cost.  The procedure has a certain probability of failing, though
we do know whether it has succeeded or not.  The failure probability 
depends on both the program and on the data state.

It was shown by Vidal and Cirac, that it is possible to increase the
probability of success by increasing the dimensionality of the
program register \cite{Vidal2000}.  They started with a single C-NOT 
processor, in their case the control qubit was the data qubit and the 
target qubit was the program qubit, that implemented the one-parameter
set of unitary operations, $U(\alpha )=\exp (i\alpha\sigma_{z})$
on the data qubit.  The probability of success is $1/2$.  By
increasing the size of the program to two qubits and adding a
Toffoli gate, they were able to increase this probability to
$3/4$.  Adding yet more qubits to the program and gates to the
processor allowed them to make the success probability as close to
one as they wished.

Another type of probabilistic quantum processor, based on the
quantum cloning circuit, was studied by us
in an earlier paper \cite{Hillery2002a}. Its qubit version (it
can be generalized to qudits) can implement any linear operator
(up to normalization) on the input qubit state.
There are still many open questions with respect to probabilistic
quantum processors, and we will study some of them in a forthcoming
publication \cite{Hillery2002b}.

\acknowledgements
We thank Matyas Koniorczyk for stimulating discussions.
This work was supported in part
by the European Union  projects EQUIP (IST-1999-11053), QUBITS
(IST-1999-13021),  QUEST (HPRN-CT-2000-00121), 
and by the National Science Foundation under grant
PHY-9970507.



\end{multicols}

\begin{thebibliography}{99}
\bibitem{blabla2}
Whether there are any nontrivial covariant processors
in the case that $\mathcal{S}={\cal H}_p$ is an open 
question.
\end{thebibliography}
\end{document}